# Analytical model of an ion cloud cooled by collisions in a Paul trap

P. Delahaye[1]

[1]GANIL, CEA/DSM-CNRS/IN2P3, Bd Henri Becquerel, 14000 Caen, France

delahaye@ganil.fr

**Abstract.** A simple model of a trapped ion cloud cooled by collisions in a buffer gas in a Paul trap is presented. It is based on the customary decomposition of the ion motion in micro- and macro- (or secular) motions and a statistical treatment of hard-sphere collisions and ion trajectories. The model also relies on the evidence that the effective trapping area in real Paul traps is limited to a certain radius, where the harmonics of the potential of order >2 become non negligible. The model yields analytical formulae for the properties of the ion cloud and equilibration times, which are in good agreement for a wide range of parameters with the results of a numerical simulation, whose reliability has been verified. When the confining potential is efficient enough to suppress evaporation from the trap, the model yields an effective temperature for the ions $T_{eff} = 2T/(1 - \frac{m_g}{m})$, where $T$ is the temperature of the buffer gas, $m$ and $m_g$ are the masses of the ions and gas molecules respectively. The so-called Radio Frequency (RF) heating effect, responsible for $T_{eff} > T$, is interpreted in light of the model as the result of an incomplete cooling of the ion motion, limited to the macromotion, while the net effect of the micromotion is to double the average ion kinetic energy for $\frac{m}{m_g} \gg 1$. For $\frac{m}{m_g} \leq 1$, the incomplete cooling is not sufficient to overcome the thermal agitation of the cloud to which the micromotion participates; the ions are therefore led out of the trap. When a thermal equilibrium is found, the dimensions of the cloud are shown to be proportional to the square root of the effective temperature: $\sigma_x = \sigma_y = \sigma_r = 2\sigma_z \propto \sqrt{T_{eff}}$. In the frame of the model, the number of collisions required to the complete cooling of the ion cloud is simply approximated by $\frac{m}{\mu} \cdot 3.5$, where $\mu$ is the reduced mass of the system. When the confining potential does not prevent evaporation from the trap, an approximate formula is derived for the evaporation rate that primarily depends on the ratio of the maximal energies of ions that can be trapped to the ion thermal energies. The comparison of the characteristic times of both processes permits to predict if the ion cloud will reach a thermal equilibrium before being evaporated.

## 1. Introduction

Ion traps are now common tools in nuclear science permitting to extend greatly the possibilities for radioactive ion beam manipulations. Among the methods derived from these devices, the buffer gas cooling in Paul traps is certainly one of the most universal and popular. In particular, the so-called RFQ beam coolers [1], are nowadays extensively used as preparation traps for precision experiments aiming at e.g. probing nuclear structure via laser spectroscopy, via high and ultra-high-accuracy mass measurements, or probing weak interaction physics in a number of more or less elaborate setups. Despite its success, the technique of buffer gas cooling in Paul traps has not yet been so far fully described in *an analytical manner*. In particular, the temperature of ion clouds in RFQ beam coolers has always been inferred from Monte Carlo simulations, despite early attempts to compare experimental measurements with theoretical studies originally done for traps used for atomic physics [2]. The latter studies [3] have been giving so far either only partial information on the cooling and ion cloud parameters [4], or considered conditions not always fulfilled in the case of RFQ coolers, such as that the ion cloud properties were governed by space charge as dominating effects [5]. Some were relying on a Langevin approach, which involved hypothesis on friction and diffusion constants as an input [6]. Lately, a study of the RF heating for 3D Paul traps permitted to derive analytical formulae relating the size of the ion cloud to the temperature, using a parametrization of the damping constant close to the transition to Coulomb crystals [7]. Again, this study applies mostly in space charge dominated regimes. The formulae have predictive powers only for very tiny cloud sizes, which are one of the necessary experimental inputs. Latest studies from chemical reactions indicate that the effective temperature of polyatomic ions is close to the one of the gas bath [8,9], but only apply to molecular ions with an internal degree of freedom [10]. For atomic ions in Paul traps, numerous experimental observations show that the ion cloud temperature is sizably larger than the one of the neutral bath, even for a low number of ions. This experimental evidence is corroborated by results from advanced simulations using realistic potentials [11]. Since the early application of the buffer gas cooling of ions in Paul trap, the so-called RF heating effect is invoked as the underlying mechanism explaining the higher temperature [12]. The aforementioned theoretical studies were so far not providing a simple temperature formula, which could allow a direct comparison with values given from simulations or with experimental observations done with RFQ beam coolers. In addition to the somewhat academic interest in the physics processes involved in the RFQ beam coolers, some precision experiments require a precise control of the trapped ion cloud parameters, which further motivate the development of such an analytical model. This is for example the case of the LPCTrap setup [13], which is being upgraded for the MORA project [14]. A Monte Carlo simulation has been undertaken to understand the present limitations of the trapping setup [15]. In this article, a simple model is developed, based on the pseudo-potential approximation, and a statistical description of hard sphere collisions. The model generally applies to collision cooling in all Paul trap devices, neglecting in a first approximation space charge effects. It sheds new light on the dynamics of the ion cooling mechanism, and provide estimates for the parameters of the ion cloud at equilibrium. Simple analytical formulae are derived for the effective temperature and dimensions of the cloud, as well as cooling times and evaporation rates. The estimates are compared to results of simulations, whose reliability has been verified for investigating experimental observations with the LPCTrap setup. The analytical formulae yield sufficiently accurate results to serve as rules of thumb for designing future Paul traps for buffer gas cooling with controlled performances.

## 2. Ion motion and collisions description

### 2.1 RF-driven ion motion description

In Paul traps, ions are confined by means of a Radio Frequency (RF) potential. In the following, we use similar notations as in [16], and consider that a pure AC (Alternative Current) potential of the form $V_{rf}(t) = V_0 \cos(\Omega t)$ is applied to one of the trap electrodes, which can be either a linear (RFQ-like) or hyperbolic (3D) Paul trap. In these conditions, one can generically describe the RF-driven ion motion in the $u$ coordinate using the Mathieu equation:

(1) $\quad \ddot{u} = -2q_u \cdot \cos(2\tau_u) \cdot u.$

$\ddot{u}$ are second derivative with respect to $\tau_u$, defined thereafter. Eq. (1) applies to hyperbolic as well as linear Paul traps in the conditions detailed below:

- In the case of a hyperbolic Paul trap, $u$ refers to the radial $x$, $y$, and axial $z$ motions. In these coordinates one defines the corresponding Mathieu parameters $q_u$:

(2) $\quad q_z = 2q_x = 2q_y = 2q_r = \frac{4qV_0}{mr_0^2\Omega^2}.$

$m$ is the mass and $q$ is the electric charge of the ion. In a cylindrical Paul trap, $r_0$ defines the radius of the ring electrode. For simplicity, we assume that the distance of the end caps to the center of the trap, $z_0$, is such that $2z_0^2 = r_0^2$. In order to outline the Mathieu equation, we used $\tau_x = \tau_y = -\tau_z = \frac{\Omega t}{2}$.

- In the case of a linear Paul trap, u refers to the radial x and y motions. In these coordinates one defines the corresponding Mathieu parameters $q_u$:

(3) $\quad q_x = q_y = q_r = \frac{2qV_0}{mr_0^2\Omega^2}.$

In a linear Paul trap, $r_0$ defines the distance of the quadrupole rods to the central axis of the trap. In order to outline the Mathieu equation, we used $\tau_x = -\tau_y = \frac{\Omega t}{2}$.

In an ideal Paul trap the criterion for ion motion stability corresponds to $q_u$<0.908. For $q_u \ll 1$, it is customary to approximate the ion motion as a sum of a rf-driven micro-motion, and a lower frequency macro- or secular motion:

(4) $\quad u \approx U + \delta.$

The macro-motion $U$ is a slow but large amplitude motion around the center of the trap:

(5) $\quad U = U_0 \cos(\omega_u t + \varphi_u)$

$U_0, \varphi_u$ are respectively an amplitude and a phase that depend on the initial conditions of the ion entering the trap. The frequency of the macro-motion is approximately given by

(6) $\quad \omega_u \approx \frac{q_u}{2\sqrt{2}}\Omega.$

The rf-driven micro-motion is a motion which is centered on the macro-motion:

(7) $\quad \delta = -\frac{q_u}{2} U \cos(\Omega t) \approx -\sqrt{2}\frac{\omega_u}{\Omega} U \cos(\Omega t).$

In these conditions one can define a pseudo-potential depth [16] which corresponds to the maximal kinetic energy which can be trapped in any given dimension. In the axial dimension and in the case of an ideal Paul trap, it is defined by

(8) $\quad D_u = \frac{q_u \times V_0}{8}.$

In the case of a real trap, the effective trapping region is limited to a given radius, $r_{eff}$. Beyond this radius, the multipoles of order greater than 2 render the ion trajectories unstable by leading the ions out of the trap, thus limiting the trapping efficiency away from the trap center (see for example [17]). The extension of the effective, n=2 harmonic trapping region depends on the geometry of the trap electrodes, which can therefore be optimized in order to minimize the importance of the unwanted multipoles. For LPCTrap, $r_{eff}$ is found to correspond to a contribution of multipoles of order n>2 of about 10% [15].

The maximum potential at the rim of the effective trapping region $r_{eff}$ can be expressed as

(9)    $V_{eff} = V_0 \left(\frac{r_{eff}}{r_0}\right)^2$.

With this in mind, we define effective pseudo-potential depths in the u-coordinate by

(10)    $D_{ueff} = D_u \left(\frac{r_{eff}}{r_0}\right)^2$.

While for traps with hyperbolic electrodes $D_{ueff}$ is expected to approach $D_u$, it is not the case for traps with more exotic geometries. For LPCTrap, the effective pseudo-potential depths correspond only to a tiny fraction of the pseudo-potential depths of an ideal trap. It was found that $\left(\frac{r_{eff}}{r_0}\right)^2 \approx \left(\frac{7}{12.9}\right)^2 = 0.29$ [15]. As it will become clear in the following, the value of $r_{eff}$ will matter for the determination of the evaporation rate. For any new trap geometry, this value would have to be determined by dedicated simulations.

### 2.2 Collisions description

In the following we describe the effect of elastic collisions of ion with the buffer molecules using a hard sphere approximation, which is also extensively used in Monte Carlo simulations. This approximation was compared to realistic potentials in [11], and has shown to provide correct estimates of the ion cloud properties, such as mean energies and radii. Reasonable agreement has also been found for the cooling and evaporation rate, when considering cross sections consistent with the mobility data [18], as discussed in Ref. [15].

In the hard sphere approximation, the elastic collision between the ion of mass $m$ and speed $\vec{v}$ and the buffer gas molecule of mass $m_g$ and speed $\vec{v}_g$ is depicted as a particle of reduced mass $\mu = \frac{m \cdot m_g}{m+m_g}$ and of relative speed $\vec{w} = \vec{v} - \vec{v}_g$ reflected by a sphere of radius $r_{hs}$ (see Fig. 1). $r_{hs}$ corresponds to the distance of closest approach of the buffer gas molecule and of the ion so that one can define the effective, geometrical, hard sphere cross section as $\sigma_{hs} = \pi \cdot r_{hs}^2$. The resulting speed $\vec{w}' = \vec{v}' - \vec{v}_g'$ can be expressed as

(11)    $\vec{w}' = \vec{w} \cdot \cos(2\theta_c)$, with $b = r \cdot \sin(\theta_c)$ the effective impact parameter.

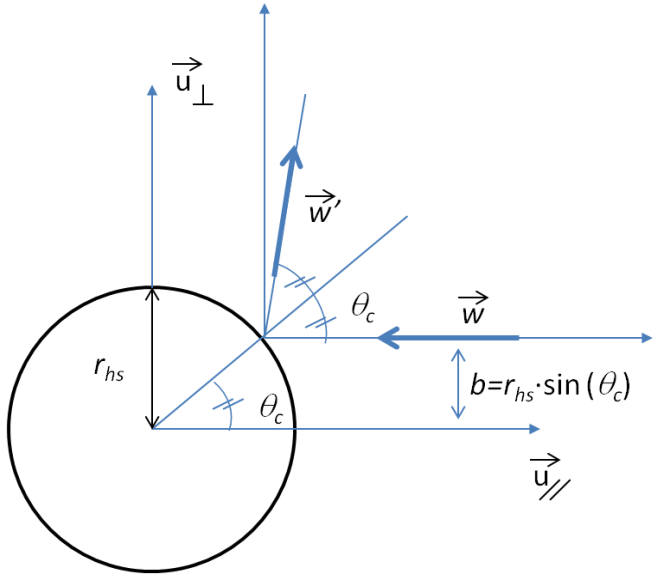

Figure 1 : Hard sphere collision parameters. See text for details.

The resulting $\vec{v}'$ speed is obtained via a frame transformation leading to

(12) $\quad \delta\vec{v} = \vec{v}' - \vec{v} = \frac{\mu}{m}(\vec{w}' - \vec{w}) = \frac{\mu}{m} \cdot \delta\vec{w}.$

We note in the following $\langle X \rangle_c$ as the average of the quantity $X$ over the collisions. Observing that for this description of the collisions, the distribution of $\sin(\theta_c)^2$ is flat, and that $\langle \vec{u}_\perp \rangle_c = \vec{0}$, one can calculate the average values

(13) $\quad \langle \delta\vec{w} \rangle_c = \langle \vec{w}' - \vec{w} \rangle_c = -\vec{w}$

And

(14) $\quad \langle \delta\vec{w}^2 \rangle_c = \langle (\vec{w}' - \vec{w})^2 \rangle_c = 2\vec{w}^2.$

Noting in addition that $\langle \vec{v}_g \rangle_c = \vec{0}$, $\langle \vec{v}_g^{\,2} \rangle_c = \frac{3kT}{m_g}$ and neglecting the weak correlation $\langle \vec{v} \cdot \vec{v}_g \rangle_c$, we obtain

(15) $\quad \langle \delta\vec{v} \rangle_c = -\frac{\mu}{m}\vec{v}$

And

(16) $\quad \langle \delta\vec{v}^2 \rangle_c = \left(\frac{\mu}{m}\right)^2 \left(2\vec{v}^2 + \frac{6kT}{m_g}\right).$

In the following collisions are described independently in each $u=x, y, z$ dimension as events leading to a velocity change which can be evaluated:

(17) $\quad \delta v_u = -\frac{\mu}{m} v_u (1 + \varepsilon).$

$\varepsilon$ follows here a centered Gaussian distribution so that $\langle\varepsilon\rangle_c = 0$ and $\langle\varepsilon^2\rangle_c = 1 + \frac{2kT}{m_g v_u^2}$ to satisfy Eq. (15) and (16). This further approximation permits a statistical treatment of the hard sphere collisions which will be particularly useful in obtaining the rate of evaporation in terms of number of collisions.

### 2.3 Average values

For the development of the model, other types of average values are calculated than those over the collisions: over the phase space of the ion cloud, or over the RF period of the micro-motion. The phase space of the ion cloud is fully determined in the pseudo-potential approximation by the secular motion phases and amplitudes. We define the following average values for the RF-driven $u$ dimension:

- $\langle X\rangle_{\varphi_u}$ is the average of the quantity $X$ over $\varphi_u$, the secular motion phase of the ions populating the cloud, assuming that $\varphi_u$ obeys a flat distribution over $[0,2\pi]$.
- $\langle X\rangle_{U_0}$ is the average of the quantity $X$ over the amplitudes of the motions of the ions populating the cloud. Assuming a Maxwell Boltzmann distribution of speeds, we show in the following section that $U_0$ obeys a centered Gaussian distribution, with $\langle U_0^2\rangle_{U_0} = \frac{kT_{eff}}{m\omega_z^2}$, and where $T_{eff}$ is the effective temperature of the ion cloud (Sec. 3.2.2).
- $\langle X\rangle_{rf}$ is the average of the quantity $X$ over the time $t$ following a flat distribution for the RF period $[0,2\pi/\Omega]$.

## 3. Properties of the ion cloud

### 3.1 Properties of individual ions

Using Eq. (4) – (7) we infer velocities, neglecting the second order terms in $\omega_u/\Omega \approx q_u/2\sqrt{2}$

(18) $\quad v_u \approx -U_0\omega_u[\sin(\omega_u t + \varphi_u) - \sqrt{2}\cos(\omega_u t + \varphi_u)\sin(\Omega t)]$.

The average energies over the RF motion then read

(19) $\quad \langle E_{ku}\rangle_{rf} \approx \frac{1}{2}m(U_0\omega_0)^2$.

We observe in inferring Eq. (18) and (19) that the micro- and macro-motion carry in average over the secular period about the same contribution to the kinetic energy, of the order of $\frac{1}{4}m(Z_0\omega_z)^2$. Formula (19) gives therefore an energy, which is about two times larger than a particle trapped in a harmonic potential whose motion would solely be determined by Eq. (5). In an ideal trap, the maximum kinetic energy of an ion that can be trapped in a given direction is limited by the maximal amplitude the trajectory can take: $X_0 = Y_0 = r_0$ and $Z_0 = z_0 = r_0/\sqrt{2}$ in the axial direction of an hyperbolic Paul trap. Using Eq. (6), one can verify that this energy corresponds to the pseudo-potential depth given in Eq. (8):

(20) $\quad \langle E_{kx}\rangle_{rf,max} = \langle E_{ky}\rangle_{rf,max} \approx \frac{1}{2}m(r_0\omega_r)^2 = qD_r = q\frac{q_r \times V_0}{8}$

And in the axial direction of an hyperbolic Paul trap:

(21) $\quad \langle E_{kz}\rangle_{rf,max} \approx \frac{1}{2}m(z_0\omega_z)^2 = qD_z = q\frac{q_z \times V_0}{8} = 2qD_r$.

From Eq. (19) one can also infer the average value of $\langle v_u^2 \rangle_{rf}$, which is related to the square of the amplitudes $U_0^2$

(22) $\quad \langle v_u^2 \rangle_{rf} = (U_0 \omega_u)^2.$

Assuming a Maxwell Boltzmann distribution of velocities, the one-dimensional $v_u$ speeds obey a centered Gaussian distribution with $\sigma_{v_u} = \sqrt{\langle v_u^2 \rangle_{U_0}} = \sqrt{\frac{kT_{eff}}{m}}$. Eq. (22) which relates speeds and amplitudes show that the amplitudes $U_0$ will also follow a Gaussian distribution of root mean square

(23) $\quad \sigma_{U_0} = \sqrt{\langle U_0^2 \rangle_{U_0}} = \frac{1}{\omega_u}\sqrt{\frac{kT_{eff}}{m}}$

And which is truncated to a maximum $U_0 \leq u_{eff}$. For the radial dimensions of both the linear and hyperbolic traps, $u_{eff} = r_{eff}$, while for the axial dimensions of the hyperbolic trap $u_{eff} = z_{eff} = \frac{r_{eff}}{\sqrt{2}}$, as we assumed $2z_0^2 = r_0^2$.

### 3.2 Properties of the ion cloud

#### 3.2.1 Master equation for inferring the properties of the cloud

In the following we define the primed variables as characteristics of the ion motion just after collision. At the time of the collision, the position of the ion remains unchanged, while the velocity changes. In the $u$ coordinate, these conditions can be expressed by means of Eq. (4) - (7) and (18) such that

(24) $\quad U_0' \cos(\omega_u t + \varphi_u') = U_0 \cos(\omega_u t + \varphi_u)$

and

(25) $\quad U_0' \omega_u \sin(\omega_u t + \varphi_u') - U_0 \omega_u \sin(\omega_u t + \varphi_u) = -\delta v_u.$

The difference of squared velocities is furthermore

(26) $\quad v_u'^2 - v_u^2 = 2 v_u \delta v_u + \delta v_u^2.$

Combining Eq. (24), (25) and (26) we find the master equation that we will use in the following to infer the properties of the ion cloud:

(27) $\quad \omega_u^2 (U_0'^2 - U_0^2) = -2 U_0 \omega_u \sin(\omega_u t + \varphi_u) \delta v_u + \delta v_u^2.$

In the hard sphere approximation for collisions, the average values of Eq. (15) and (16) for $\langle \delta \vec{v} \rangle_c$ and $\langle \delta \vec{v}^2 \rangle_c$ can be used.

#### 3.2.2 Temperature of the ion cloud

We will use in the following Eq. (27) with different averages, on the phase space of the ion cloud, RF motion and collisions in order to deduce the temperature of the ion cloud, assuming that a thermal equilibrium is found between the ions and the molecules of the buffer gas. At equilibrium, the averaged term on the left hand side of Eq. (27) is expected to cancel out as the average amplitudes after and before the collisions should tend to equalize. The average on the RF motion is important to get rid of the ion cloud size rapid beating due to the micro-motion excitation. One expects to find an effective temperature $T_{eff} > T$, because of the well known rf heating effect [12]. To our knowledge, such effect has never been fully self-consistently estimated other than by numerical simulations, which were widely used for simulating RFQ cooler bunchers (see for instance [19-22]). So far models [2-7] have been deriving information on the RF heating, from other experimental parameters such as e.g. the ion damping constant, the ion cloud size, and number of ions in case of space charge dominated regimes.

Using Eq. (15) and (16) to estimate an average of Eq. (27) over collisions one obtains

(28) $\quad \omega_u^2 \langle (U_0'^2 - U_0^2) \rangle_c = 2 \frac{\mu}{m} v_u U_0 \omega_u \sin(\omega_u t + \varphi_u) + 2 \left(\frac{\mu}{m}\right)^2 (v_u^2 + \frac{kT}{m_g})$.

Using Eq. (18) for $v_u$, and averaging over the RF cycle one obtains

(29) $\quad \omega_u^2 \langle (U_0'^2 - U_0^2) \rangle_{c,rf} = \left(\frac{\mu}{m}\right)^2 \left(2U_0^2 \omega_z^2 + \frac{2kT}{m_g}\right) - \frac{\mu}{m} \cdot 2U_0^2 \omega_u^2 \sin(\omega_u t + \varphi_u)^2$.

Averaging further on the ion cloud phase space one should obtain, assuming an ion cloud at equilibrium, no change of average amplitudes so that

(30) $\quad \omega_u^2 \langle (U_0'^2 - U_0^2) \rangle_{c,rf,\varphi_z,U_0} = 0 = \left(\frac{\mu}{m}\right)^2 \left[2\langle U_0^2 \rangle_{v_u} \omega_u^2 \left(1 - 1/2 \cdot \frac{m}{\mu}\right) + \frac{2kT}{m_g}\right]$.

From Eq. (23), relating $\langle U_0^2 \rangle_{U_0}$ to $T_{eff}$, and Eq. (30), one deduces the effective temperature of the ion cloud:

(31) $\quad T_{eff} = \frac{2T}{(1 - \frac{m_g}{m})}$.

As stated above, Eq. (31) only holds in the pseudo-potential approximation limit, and in the case of an ion cloud in equilibrium, when the evaporation is negligible. We will show in the following that the rate of evaporation depends mostly of the ratios $\alpha_{Er} = \frac{D_{reff}}{kT_{eff}}$, and $\alpha_{Ez} = \frac{D_{zeff}}{kT_{eff}}$ in the case of the 3D Paul trap. As we based our reasoning on a generic coordinate *u*, all RF-driven motions will tend to reach the same temperature as in Eq. (31) for the domain of validity of the model. The temperature does not depend on the Mathieu parameter. As a result, and despite the asymmetry of confining potentials in radial and axial dimensions, the equipartition in energy also holds for ions trapped and cooled by buffer gas in a 3D Paul trap. In contrast, in linear Paul traps such as RFQ coolers, only the radial *x*, *y* dimensions are experiencing the RF heating such that their temperature reaches Eq. (31). Diffusion laws drive the axial motion in the static potential of the trap [18]. When confined in a potential well, as it is the case in RFQ cooler bunchers, the *z* motion is in thermal equilibrium at the temperature of the buffer gas.

The accuracy of Eq. (31) for the RF-driven motions in the linear and 3D Paul traps has been probed for several parameters in Sec. 4. For $\frac{m}{m_g} \gg 1$, the effective temperature is roughly equal to 2 times the temperature of the gas. This factor was experimentally observed with different devices, as could be reported for instance in the early development of RFQ coolers as was done at Mc Gill university [2, 19, 22] for a low number of ions, and has lately been shown to be in good agreement with LPCTrap data [15].

Eq. (31) has been obtained by averaging of the RF period. In this respect it is worth noticing that the instantaneous average energy of the ions over the cloud phase space oscillates at twice the RF frequency around the expected thermal energy. Using Eq. (18) and averaging the energy over the ion cloud phase space, instead of the RF period as was done to obtain Eq. (19), one obtains

$$(32) \quad \langle E_{ku} \rangle_{\varphi_u, U_0} \approx \frac{1}{4} m \langle U_0^2 \rangle_{U_0} \omega_u^2 (1 + 2\sin(\Omega t)^2) = 1/2 \, kT_{eff}(1 - \cos(2\Omega t)/2).$$

Eq. (32) reproduces quite accurately the beating that can be observed using the simulations described in Sec. 4.

Finally it is worth commenting the derivation of the temperature which sheds a new light on the origin of RF heating effect [12], by emphasizing the role of the secular motion phase in Eq. (29). In contrast with what would look like the standard description of a simple Brownian motion of a free particle, the term damping the velocity ($2v_u \delta v_u$ in Eq. (26)) is weighted by $\sin(\omega_u t + \varphi_u)^2$. The latter term, when averaged over the secular phase, generates the factor of 2 which approximately relates the effective temperature of the ions to the one of the gas for $\frac{m}{m_g} \gg 1$ in Eq. (31). The energy dissipation is therefore limited to the secular motion, which enters in thermal equilibrium with the gas, while the RF motion, coupled to the secular motion via Eq. (7), acts as an additional degree of freedom whose net effect is to double the kinetic energy (see discussion in Sec. 3.1). When the ratio $\frac{m}{m_g}$ becomes smaller than 1, i.e. $\frac{m}{\mu} \leq 2$ in Eq. (30), this incomplete cooling is not sufficient to prevent the thermal agitation term ($\delta v_u^2$ in Eq. (26)) to which the micro-motion participates to lead ions out of the trap. Compared to the Brownian motion, the cooling in the RF trap therefore appears to be less efficient, incomplete, as it only affects the macromotion, while the micromotion participates to the thermal agitation by doubling the energy of the ions for $\frac{m}{m_g} \gg 1$, eventually leading ions out of the trap for ratios $\frac{m}{m_g} \leq 1$.

### 3.2.3 Dimensions of the ion cloud

As discussed in Sec. 3.1, we assumed a Maxwell Boltzmann distribution of velocities, which is equivalent in the pseudo-potential approximation to assume Gaussian distributions for the amplitudes of the secular motion. Eq. (4) to (7) show that on average over the RF period, the ion motion is centered on the secular motion of amplitude $U_0$. From Eq. (5) the spatial widths of the ion cloud for the RF driven dimensions are:

$$(33) \quad \sigma_u^2 = \sigma_{U_0}^2 \langle \cos(\omega_u t + \varphi_u)^2 \rangle_{\varphi_u} = \sigma_{U_0}^2 / 2.$$

Averaging Eq. (19) over the ion cloud, one relates the dimensions of the cloud to the average energies of the ions of the cloud:

$$(34) \quad \langle E_{ku} \rangle_{rf, \varphi_u, U_0} \approx \frac{1}{2} m \langle U_0^2 \rangle_{U_0} \omega_u^2 = m \omega_u^2 \sigma_u^2.$$

Consistent relationships between the ion cloud radius and energy can be derived from [3] in the frame of the so-called "Modified pseudo-potential Model", and from [7] when neglecting space charge effects (i.e. setting the number of ions N=0). Recalling Eq.(20) and (21) one gets, for the radial dimensions

$$(35) \quad \sigma_x^2 = \sigma_y^2 = \sigma_r^2 = \frac{\langle E_{kx} \rangle_{rf, \varphi_x, X_0}}{m \omega_r^2} = \frac{\langle E_{kx} \rangle_{rf, \varphi_x, X_0}}{2qD_r} r_0^2 = \frac{\langle E_{ky} \rangle_{rf, \varphi_y, Y_0}}{2qD_r} r_0^2$$

And for the axial dimension of a 3D Paul trap

(36) $\quad \sigma_z^2 = {\sigma_{Z_0}^2}/{2} = \frac{\langle E_{kz}\rangle_{rf,\varphi_z,Z_0}}{m\omega_z^2} = \frac{\langle E_{kz}\rangle_{rf,\varphi_z,Z_0}}{2qD_z} Z_0^2.$

Using the temperature derived in Sec. 3.2.2 for the RF-driven motions, the model presented here gives the following predictions for the dimensions of the ion cloud:

(37) $\quad \sigma_x^2 = \sigma_y^2 = \sigma_r^2 = \frac{kT_{eff}}{4qD_r} r_0^2 = \frac{2}{(1-\frac{mg}{m})} \cdot \frac{kT}{4qD_r} r_0^2$

and in the case of a 3D Paul trap

(38) $\quad \sigma_z^2 = \frac{2}{(1-\frac{mg}{m})} \cdot \frac{kT}{4qD_z} Z_0^2 = \frac{1}{4}\sigma_r^2.$

For the latter trap, the spatial extension of the cloud in the *z* axis is therefore 2 times smaller than in the radial dimension. One deduces the average spherical squared radius in the hyperbolic trap summing contributions from Eq. (35) and (36):

(39) $\quad \langle \rho^2 \rangle_{rf,X_0,\varphi_x\ldots} = \sigma_x^2 + \sigma_y^2 + \sigma_z^2 = \frac{3}{8} \frac{\langle E_k \rangle_{rf,cloud}}{qD_r} r_0^2.$

Eq. (37) can be recast to deduce the mean kinetic energy of ions from measured squared radii:

(40) $\quad \langle E_k \rangle_{rf,cloud} \approx \frac{8eD_r}{3} \cdot \frac{\langle \rho^2 \rangle_{rf,cloud}}{r_0^2} = 6eD_r \cdot \frac{\sigma_r^2}{r_0^2}.$

In the frame of the model presented here, the ion cloud is therefore approximated by Gaussian distributions for all RF-driven coordinates, with the respective width given by Eq. (37) and (38). This approximation has been found very close to what is obtained by the numerical simulation for the studied systems. A recent study shows, however, that for a ion mass approaching a sizeable fraction of the one of the buffer gas, the ion spatial distribution progressively departs from the Gaussian law. The tail distribution becomes a power law of the distance to the center of the trap, while the overall distribution maintains a seemingly constant spread [23]. The possible implications on the model predictions are explored in Sec. 4.

### 3.2.4 Characteristic cooling time

The cooling time is defined here as the number of collisions to cool down the velocities of a hot ion cloud with an initial temperature well beyond the equilibrium temperature of Eq. (31) to half speeds. From Eq. (15), one can readily define the characteristic cooling time:

(41) $\quad n_{1/2} \approx \frac{m}{\mu} \ln(2).$

One considers that a complete cooling is achieved after about 5 $n_{1/2}$, which is found in good agreement with the simulations shown in Sec. 4.

### 3.2.5 Evaporation from the ion cloud

As stated before, Eq. (31) only holds for an equilibrated ion cloud. It is interesting to evaluate what is the evaporation rate of an ion cloud to define a limit for which one can safely assume that the ion cloud will be in thermal equilibrium. In the following, we evaluate the rate of evaporation from our master equation, Eq. (27), where we describe statistically the collision events by means of Eq. (17). Using these equations and Eq. (18) for the velocity one finds

(42) $\frac{U_0'^2}{U_0^2} = 1 + 2\frac{\mu}{m}(1+\varepsilon)\left(\sin(\omega_u t + \varphi_u) - \sqrt{2}\cos(\omega_u t + \varphi_u)\sin(\Omega t)\right)\sin(\omega_u t + \varphi_u) +$
$\left[\frac{\mu}{m}(1+\varepsilon)\right]^2 \left(\sin(\omega_u t + \varphi_u) - \sqrt{2}\cos(\omega_u t + \varphi_u)\sin(\Omega t)\right)^2$

With $\varepsilon$ as defined above, with $\langle\varepsilon\rangle_c = 0$ and $\langle\varepsilon^2\rangle_c = \sigma_\varepsilon^2 = 1 + \frac{2kT}{m_g v_u^2}$.

If one averages Eq. (42) over the phase space of the cloud and RF period, it is interesting to note that at equilibrium with $U_0'^2 = U_0^2$ one finds

(43) $\langle\varepsilon^2\rangle_{c,rf,\varphi_u,\omega_u} = \frac{m}{m_g}$.

One finds again an equilibrium temperature which is consistent with Eq. (31) from the definition of $\varepsilon$, which has to comply with Eq. (15) and (16):

(44) $\langle v_u^2\rangle_{c,rf,\varphi_u,\omega_u} = \frac{2kT}{m_g(1+\langle\varepsilon^2\rangle_{c,rf,\varphi_u,\omega_u})} = \frac{2kT}{m(1-\frac{m_g}{m})}$.

The approximation of Eq. (17) yielding Eq. (42) is therefore consistent with the results obtained so far and is believed in this respect to give reasonable results.

Eq. (42) permits to describe statistically the change of amplitude of the secular motion of ions consecutive to a collision. In the model and in coherence with what was observed for instance in [15], ions are evaporated, i.e. lost after a collision if their new amplitude exceeds the dimensions of the effective trapping area:

(45) $\frac{U_0'^2}{U_0^2} > \frac{u_{eff}^2}{U_0^2} = 1 + \delta_{U_0^2}$

where one defines $\delta_{U_0^2} = \frac{u_{eff}^2 - U_0^2}{U_0^2} > 0$.

To simplify Eq. (42), as we are mainly interested in the frequency of occurrence of the condition of Eq. (45), we particularize different cases for the RF and secular motion phases with different weight factors to preserve some of the most relevant averages of the trigonometric functions. For the RF phases, we particularize $\Omega t = \pm\frac{\pi}{4}$ with equal 1/2 weight factors such that $\langle\sin(\Omega t)\rangle = 0$ and $\langle\sin(\Omega t)^2\rangle = 1/2$. This permits to rewrite Eq. (42) in the following way, using $\epsilon = \frac{\mu}{m}(1+\varepsilon)$:

(46) $\frac{U_0'^2}{U_0^2} = 1 - \epsilon + \epsilon^2 + \epsilon\sqrt{1+(1-\epsilon)^2}\cos(2(\omega_u t + \varphi_u) - \varphi_{\epsilon\pm})$.

Where $\varphi_{\epsilon\pm}$ are phases corresponding to $\Omega t = \pm\frac{\pi}{4}$ such that

(47) $\cos(\varphi_{\epsilon\pm}) = \frac{1}{\sqrt{1+(1-\epsilon)^2}}$ and $\sin(\varphi_{\epsilon\pm}) = \frac{\pm(1-\epsilon)}{\sqrt{1+(1-\epsilon)^2}}$.

We further simplify Eq. (46) by particularizing 3 secular motion phases, corresponding to $\cos(2(\omega_u t + \varphi_u) - \varphi_{\epsilon\pm}) = 0, \pm 1$ with respective weights of $1 - 2/\pi$ and $1/\pi$ for each sign to yield the following averages: $\langle\cos(2(\omega_u t + \varphi_u) - \varphi_{\epsilon\pm})\rangle = 0$ and $\langle|\cos(2(\omega_u t + \varphi_u) - \varphi_{\epsilon\pm})|\rangle = \frac{2}{\pi}$. We deduce constraints on the variable $\epsilon = \frac{\mu}{m}(1+\varepsilon)$ to satisfy Eq. (46):

(48) $\epsilon = \frac{\mu}{m}(1+\varepsilon) > \sqrt{\delta_{U_0^2}}$ for $\cos(2(\omega_u t + \varphi_u) - \varphi_{\epsilon\pm}) = 1$

(49) $\epsilon < \frac{1}{\sqrt{2}}\left(1 - \sqrt{1 - 2\sqrt{2}\,\delta_{U_0^2}/{1+\sqrt{2}}}\right)$ for $\cos(2(\omega_u t + \varphi_u) - \varphi_{\epsilon\pm}) = -1$

(50) $\epsilon < \frac{1}{2}\left(1 - \sqrt{1 + 4\delta_{U_0^2}}\right)$ and $\epsilon > \frac{1}{2}\left(1 + \sqrt{1 + 4\delta_{U_0^2}}\right)$ for $\cos(2(\omega_u t + \varphi_u) - \varphi_{\epsilon\pm}) = 0$.

In Eq. (48) and (49) we used approximate solutions based on a Taylor expansion around different $\epsilon_0$ to select the best function representing Eq. (46) for $\epsilon$ on the relevant domain. For Eq. (50) we used the exact solutions. Eq. (48) – (50) taking the form of $\epsilon > f(\delta_{U_0^2})$ or $\epsilon < f(\delta_{U_0^2})$ we deduce probabilities for evaporative collisions at a given $\delta_{U_0^2}$ by defining the centered and reduced Gaussian variable:

(51) $\tau = \frac{1}{\sigma_\varepsilon}\left(\frac{m}{\mu}\epsilon - 1\right)$.

Defining the ratios:

(52) $\alpha_\mu = \frac{m}{\mu}$

and

(53) $\alpha_{Eu} = \frac{\langle E_{ku}\rangle_{rf,max}}{kT_{eff}} = \frac{qD_{ueff}}{kT_{eff}} = \frac{m\omega_u^2 u_{eff}^2}{2kT_{eff}}$

One can rewrite $\sigma_\varepsilon$ such that:

(54) $\sigma_\varepsilon^2 \approx \langle \sigma_\varepsilon^2\rangle_{rf} = 1 + \frac{2kT}{m_g U_0^2 \omega_u^2} = 1 + \left(\frac{\alpha_\mu - 2}{2\alpha_{Eu}}\right)\cdot(1 + \delta_{U_0^2})$.

The probability of an evaporative collision at a given $\delta_{U_0^2}$ is then numerically evaluated for each secular motion phase. In the case of $\cos(2(\omega_u t + \varphi_u) - \varphi_{\epsilon\pm}) = 1$ for instance we calculate the probability of evaporation the following way:

(55) $P_{ev,\delta_{U_0^2},\cos(2(\omega_u t+\varphi_u)-\varphi_{\epsilon\pm})=1} = \frac{1}{2}\left(1 - \mathrm{Erf}\left[\frac{\frac{1}{\sigma_\varepsilon}\left(\alpha_\mu\sqrt{\delta_{U_0^2}}-1\right)}{\sqrt{2}}\right]\right)$.

The total probability of evaporative collision at given $\delta_{U_0^2}$ can be approximated as:

(56) $P_{ev,\delta_{U_0^2}} = \frac{1}{\pi}\left(P_{ev,\delta_{U_0^2},\cos(2(\omega_u t+\varphi_u)-\varphi_{\epsilon\pm})=1} + P_{ev,\delta_{U_0^2},\cos(2(\omega_u t+\varphi_u)-\varphi_{\epsilon\pm})=1} + (\pi - 2)\cdot P_{ev,\delta_{U_0^2},\cos(2(\omega_u t+\varphi_u)-\varphi_{\epsilon\pm})=1}\right)$.

This function is finally numerically convoluted with the Gaussian distribution of the ion cloud in $U_0$ (see Eq. (22) and (23)) which is steaming out from the Maxwell Boltzmann distribution in speed, and limited to $U_0^2 = u_{eff}^2$:

(57) $P_{\delta_{U_0^2}} \propto e^{-\alpha_{Eu} U_0^2}$.

We call in the following $P_{ev,u}(\alpha_\mu, \alpha_{Eu})$ the resulting function, which approximates the probability of evaporation of an ion in the RF-driven $u$ dimension over the entire cloud after a collision. $P_{ev,u}(\alpha_\mu, \alpha_{Eu})$ has been fitted using Mathematica [24] on a wide range of $\alpha_\mu, \alpha_{Eu}$ parameters. For $\alpha_\mu \in [2, 10]$ and $\alpha_{Eu} \in [0.25, 10]$, we find a function approximating very well the numerical function:

$$(58) \quad P_{ev,u}(\alpha_\mu, \alpha_{Eu}) \approx 0.382 \left(1 - \frac{1.14}{\alpha_{Eu}}\right) e^{-\alpha_{Eu}/2 - \alpha_\mu} + 0.134 \left(1 + \frac{1}{2\alpha_{Eu}}\right) \frac{e^{-\alpha_{Eu}}}{\sqrt{\alpha_\mu}}.$$

For each kind of trap, the complete probability of evaporation after a collision is approximated by summing formula (58) for the RF-driven motions. Reminding the values $u_{eff}$ defined in Sec. 3.1, $u_{eff} = r_{eff}$ for the radial dimensions of the linear and hyperbolic Paul traps, and $u_{eff} = z_{eff} = \frac{r_{eff}}{\sqrt{2}}$ for the hyperbolic trap, one gets using Eq. (53):

$$(59) \quad \alpha_{Ex} = \alpha_{Ey} = \alpha_{Er} = \frac{m\omega_r^2 r_{eff}^2}{2kT_{eff}} = \frac{\alpha_{Ez}}{2}.$$

The resulting complete probability of evaporation is

$$(60) \quad P_{ev} = P_{ev,u}(\alpha_\mu, \alpha_{Ez}) + 2 \cdot P_{ev,u}(\alpha_\mu, \alpha_{Er}), \text{ for the 3D trap.}$$

In Eq. (60), the contribution of the radial losses dominates, as the pseudo-potential depth is twice lower than for the axial dimension in such kind trap. Assuming no losses in the axial direction, one gets

$$(61) \quad P_{ev} = 2 \cdot P_{ev,u}(\alpha_\mu, \alpha_{Er}), \text{ for the linear Paul trap.}$$

Depending on the kind of trap, one deduces from Eq. (60) or (61), the average number of collisions in the hard sphere approximation, which are required to evaporate one ion from the cloud:

$$(62) \quad n_{ev} = 1/P_{ev}.$$

Eq. (62) has found to be reproducing very well the average number of collisions before evaporation in the numerical simulation for both kind of traps, as will be shown in the following. The equilibration of the ion cloud depends on the balance between cooling and evaporation, whose characteristic times are given by Eq. (41) and (62). Estimating that a complete cooling will approximately require $5\, n_{1/2}$ collisions, one can express the following condition for an equilibrated ion cloud:

$$(63) \quad \alpha_{eq} \approx \frac{n_{ev}}{5 n_{1/2}} \approx 0.3 \frac{n_{ev}}{\alpha_\mu} > 1$$

Where $P_{ev}$ is evaluated using Eq. (60) or (61) depending on the trap. This condition was verified again in the simulations, where it was observed that in practice, the condition of Eq. (63) is not satisfied for relatively low $\alpha_{Er}$ values ($\alpha_{Er} \lesssim 1$). For these values, an effective temperature smaller than predicted by the model in Eq. (31) was derived from the average kinetic energy of trapped ions.

## 4. Domain of validity of the model

The model presented here relies on several approximations, which will affect its domain of application:

- The pseudo-potential approximation [16] is valid for small Mathieu parameters.

- The spatial distribution of the ion cloud has been approximated as Gaussian, while for high $\frac{m_g}{m}$ ratios, approaching unity, the tail of the ion cloud is known to follow a power law [23].
- The static properties of the cloud (temperature, sizes), are derived assuming that the ion cloud is equilibrated (condition of Eq. (63) is satisfied).
- Space charge effects have been neglected.

In the following sections, the domain of validity of the model is explored in absence of space charge effects by comparing the model predictions with results from a Monte Carlo simulation, whose reliability was verified for investigating the measured performances of LPCTrap [15]. As test cases, one considers the cooling of $^{39}K^+$ ions with different buffer gases ($H_2$, He, and Ne) at different temperatures (4K, 77K, and 300K) in a Paul trap identical to LPCTrap, and in a linear cooler of same $r_0$ (12,875 mm). Mathieu parameters are scanned on the whole stability diagram by varying the RF voltage amplitude while maintaining the frequency to 600kHz. One finally discusses in which conditions one can safely neglect space charge effects.

### 4.1 Temperature of the ion cloud

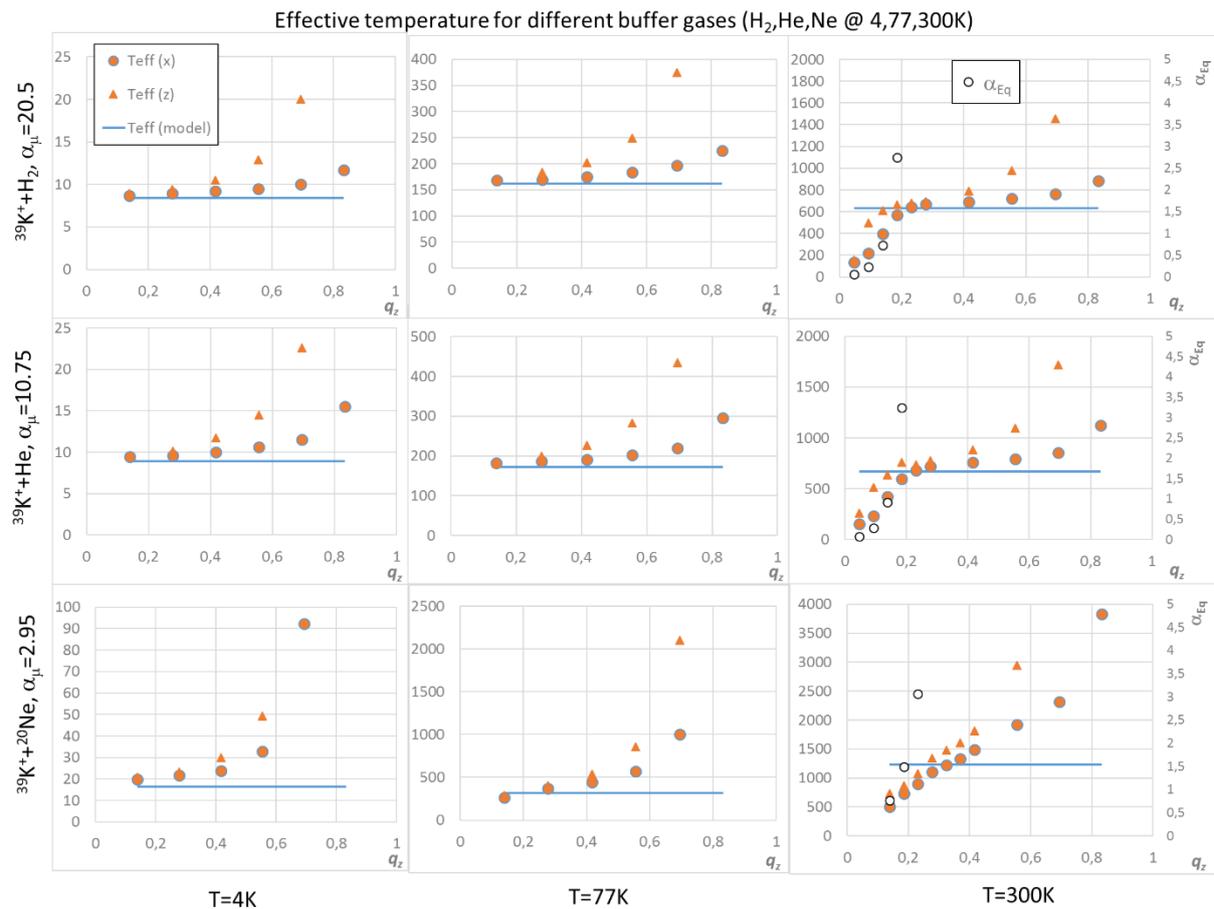

Figure 2 : Effective temperatures as derived by the model (Eq. (31)) and from results of the simulations for the x and z dimensions of the 3D Paul trap.

The model prediction for the effective temperature of RF-driven motions (Eq. (31)) is compared to simulation results in Fig. 2, for the radial and axial motions of $^{39}K^+$ ions trapped in the 3D Paul trap. The 3D Paul trap, identical to LPCTrap, presents an effective trapping radius of 7 mm, as discussed in [15]. As expected, the agreement of the model with the simulation is excellent for low $q_z$ values, and for an equilibrated cloud with $\alpha_{eq}>1$ (Eq. (63)). The contribution of harmonics of the ion motion of higher order than those considered in the frame of the pseudo-potential approximation are increasingly important for higher and higher $q_z$ values. Their contribution result in a hotter effective cloud temperature than predicted. Nevertheless, the prediction of Eq. (31) remains valid within 30% error for the axial direction and 15% error for the radial motion up to $q_z \sim 0.5$. For the whole diagram of stability, the prediction is generally better for the radial direction than for the axial direction, as $q_r = q_z/2$.

For T=300K, the condition $\alpha_{eq}>1$ is not fulfilled for low $q_z$ values, and the average energy of trapped ions in the non equilibrated cloud yield an estimate for the temperature which is lower than Eq. (31).

The dependence of $T_{eff}$ on the ratio $\frac{m_g}{m} = \frac{1}{\alpha_\mu - 1}$ is quite well reproduced by the model as shown in Fig. 3, down to $\frac{m_g}{m} \sim \frac{1}{2}$ (equivalently $\alpha_\mu \sim 3$). For a radial Mathieu parameter $q_r$=0.07, simulations and model predictions agree within better than 5%.

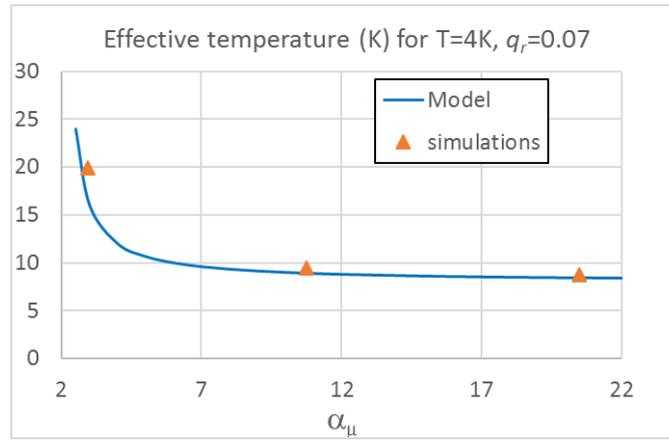

Figure 3 : Effective temperature for the radial motion as a function of $\alpha_\mu$, for T=4K and $q_r$=0.07.

The validity of Eq. (31) is verified for the radial motions of ions in a linear trap in Fig. 4 for T=77K, where it was assumed that ions were axially confined by an harmonic potential. As expected, the axial motion is at the temperature of the buffer gas, while the radial motion temperature exhibits a similar behaviour as for the 3D Paul trap. It is only at the rim of the diagram of stability ($q_r$>0.8), that the axial motion becomes hotter, influenced by the highly energetic radial motions.

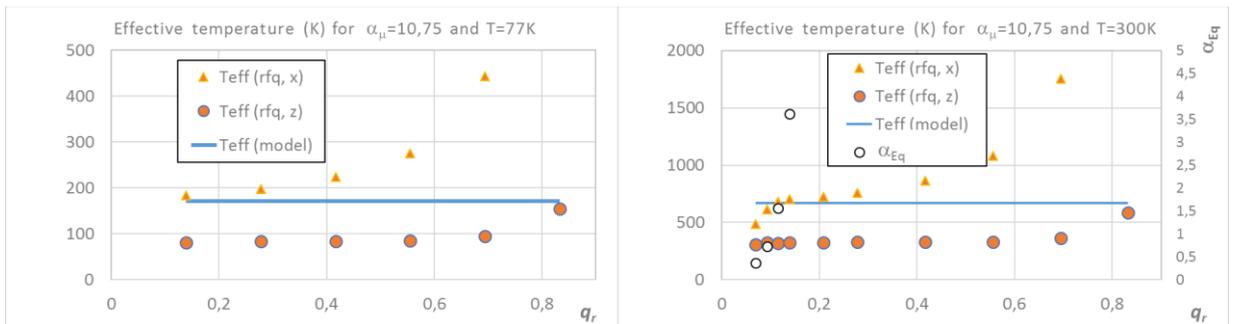

Figure 4 : Effective temperature for the radial motion and temperature of the axial motion in a linear trap, cooled by a He buffer gas at 77K.

### 4.2 Dimensions of the ion cloud

The size of the ion cloud in the different directions for the 3D Paul trap as deduced from Eq. (37) and Eq. (38) are compared to results of the simulation in Fig. 5, for ions cooled in He and Ne buffer gases at T=77K. The prediction of the model are found in good agreement with the simulation results for He. The accuracy of the prediction for $\sigma_z^2$ remains within better than 30% up to $q_z \sim 0.5$, as for the temperature. The agreement is more modest for Ne, for which a deviation of 60% of the model from the simulated value is observed at $q_z = 0.42$. In this case, the greater deviation is attributed to the limit of the Gaussian shape approximation for $\frac{m_g}{m}$ ratios approaching unity [23].

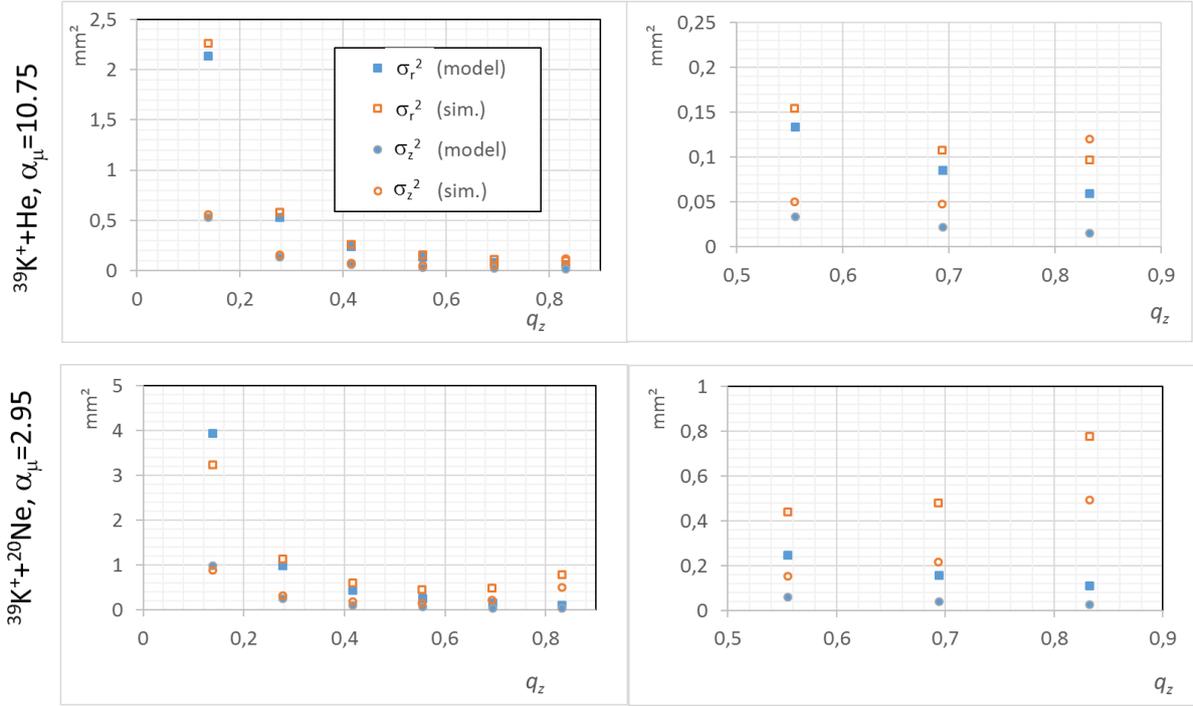

Figure 5 : Dimensions of the cloud in the 3D Paul trap for $^{39}K^+$ ions cooled in a He buffer gas at T=77K. The right insets are a zoom on the region $q_z > 0.5$.

### 4.3 Cooling time

The cooling time estimate of Eq. (41) has been compared to results of the simulations by fitting the evolution of the velocity amplitudes of an initially hot ion cloud ($\sigma_r^2$, $\sigma_z^2$) cooled in the 3D Paul trap by $H_2$, He and $^{20}$Ne buffer gases at 4K, 77K and 300K. The results of this comparison are summarized in Fig. 6, where the characteristic cooling time, $n_{1/2}$, is expressed in number of collisions. Without much surprise, the overall agreement is good, as both the model and simulation used here are based on the hard sphere approximation. As shown in [15], $n_{1/2}$ can however be reasonably well translated into cooling times by using a simple expression for the geometric collision cross section of the hard sphere approximation, discussed in Sec. 2.2:

$$(64) \quad \sigma_{hs} \approx \pi(r_w + r_i)^2$$

Where $r_w$ is the Van der Waals radius of the buffer gas atoms or molecules, and $r_i$ is the ion radius. Such approximation has shown to compare reasonably well with a simulation using realistic potentials. In these conditions, the average time between two collisions, $\tau_c$, can be approximated by

$$(65) \quad \tau_c = \frac{1}{N° \cdot \sigma_{hs} \cdot w}$$

Where $N°$ is the number of buffer gas atoms/molecules per volume unit, which depends on the buffer gas pressure and temperature, and $w$ is the average speed between the ions and the gas molecule. For the latter, one can use the estimate:

$$(66) \quad w \approx \sqrt{\frac{kT}{m_g} + \frac{kT_{eff}}{m}}$$

Where $T_{eff}$ is given by Eq. (31).

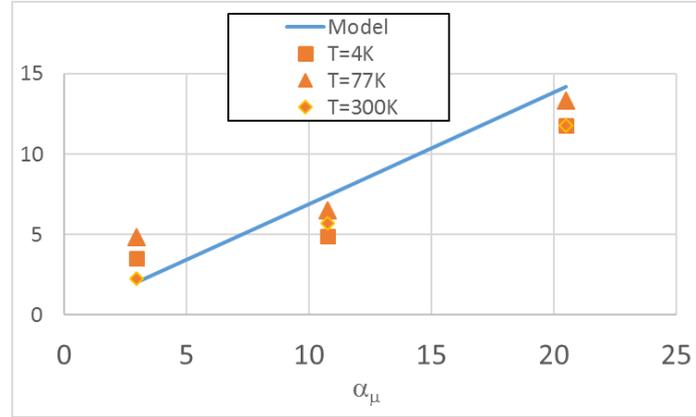

Figure 6 : characteristic cooling times expressed in number of collisions for different H$_2$, He, and Ne buffer gases at 4K, 77K and 300K. For H$_2$, $\alpha_\mu$=20.5, for He, $\alpha_\mu$=10.75 and for Ne, $\alpha_\mu$=2.95.

### 4.4 Evaporation time

The estimate for the average number of collisions before evaporation, $n_{ev}$, given in Eq. (62) is compared to results of the simulations in Fig. 7 for the 3D Paul trap, and in Fig. 8 for the linear Paul trap. In the latter, one assumes again that ions are trapped in the axial direction by an electrostatic harmonic potential. The effective trapping radius is set artificially to 7 mm, i.e. identical to the 3D Paul trap: in the simulation, ions are considered lost as soon as their orbit goes beyond this limit. The comparison is done for T=300K, for which the evaporation, primarily driven by the ratio $\alpha_{Er} = \frac{qD_{reff}}{kT_{eff}}$, is the most important. For equivalent Matthieu parameters, the confining voltages of the linear Paul trap are twice larger, yielding pseudo-potential depths twice larger than in the case of the 3D Paul trap. Equivalent $n_{ev}$ values are therefore found for Mathieu parameters twice lower than in the case of the 3D Paul trap. For both ion traps, one observes a stunning agreement of the predicted evaporation lifetimes with the results of the simulations for Mathieu parameter values below or equal to 0.3. The agreement is almost as good for Ne as for He as buffer gases, showing that the approximation of a Gaussian cloud yield still accurate predictions for mass ratios as large as $\frac{m_g}{m} \sim \frac{1}{2}$. As can be seen for the 3D Paul trap with Ne as buffer gas, beyond $q_z = 0.3$, the harmonic motions of higher order than those considered in the pseudo-potential approximation yield an RF heating resulting in evaporation losses not accounted for in the model. As for $n_{1/2}$, the average number of collisions before evaporation $n_{ev}$ can be translated into an average evaporation time for given experimental conditions (buffer gas pressure and temperature) by using the simple approximations discussed in Sec. 4.3 (Eq. (65) -(67)).

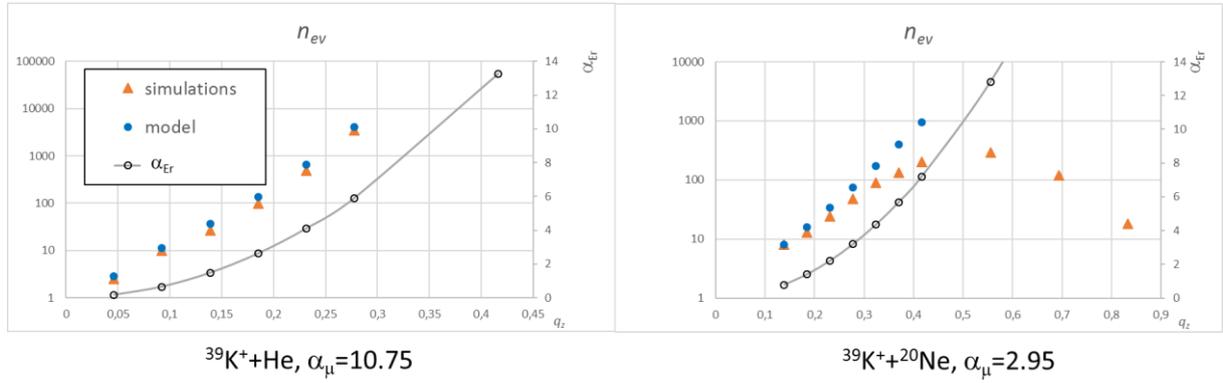

$^{39}$K$^+$+He, $\alpha_\mu$=10.75     $^{39}$K$^+$+$^{20}$Ne, $\alpha_\mu$=2.95

**Figure 7**: average number of collision before evaporation for the 3D Paul trap with He and Ne buffer gases, at T=300K.

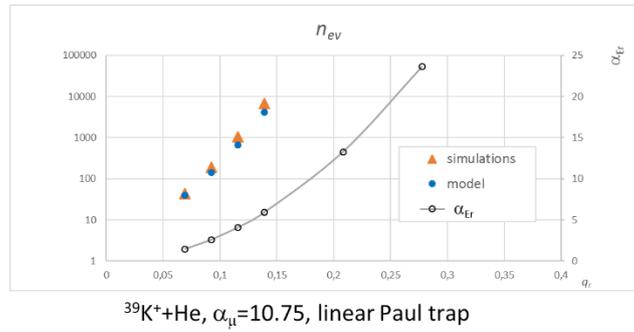

$^{39}$K$^+$+He, $\alpha_\mu$=10.75, linear Paul trap

**Figure 8**: average number of collisions before evaporation for the linear Paul trap for He as buffer gas and T=300K.

### 4.5 Space charge considerations

The model presented here neglects space charge effects. In practice, the dimensions of the cloud derived in its framework permit to estimate until what number of trapped ions this assumption will hold. For example, for the 3D Paul trap, one can express the maximal charge density in the center of the trap $\rho_0$ as a function of the number of trapped ions $N$:

$$(69) \qquad \rho_0 = \frac{qN}{(2\pi)^{\frac{3}{2}} \cdot \sigma_r^2 \cdot \sigma_z}.$$

In order to neglect safely the space charge effects in the Paul trap, $\rho_0$ has to have a negligible impact on the confining potential. Using the Poisson law, one can derive from the pseudo-potential approximation the maximal charge density that the trap can hold:

$$(70) \qquad \rho_{max} = \frac{12\varepsilon_0 D_{reff}}{r_{eff}^2}.$$

One can deduce from Eq. (37) and (38) that one can safely neglect space charge effects for ion numbers satisfying

$$(71) \qquad N \ll 3\left(\frac{\pi}{\alpha_{Er}}\right)^{\frac{3}{2}} \cdot \varepsilon_0 \cdot D_{reff} \cdot r_{eff}.$$

Taking the example of LPCTrap, one can estimate that Eq. (71) is satisfied up to $N \sim 10^4$ using a buffer gas at room temperature, while for cryogenic temperatures limits are shrinking because of the small dimensions of the cloud: one get $N \sim 10^3$ for 77K, and $N \sim 10^2$ for 4K.

Given some assumptions on the axial confining potentials, a similar reasoning can easily be undertaken for linear traps to derive an equation similar to Eq. (71).

## 5. Conclusion

The model presented here permits to infer properties of an ion cloud trapped and cooled by buffer gas in a Paul trap in the absence of space charge effects. It sheds new light on the RF heating effect. This effect has never been quantified so far in a model in a self-consistent manner, without the input of experimental observables such as the size of the ion cloud. It provides an estimate for the effective equilibrium temperature of the ions for the RF-driven motions, $T_{eff} > T$, when the evaporation from the cloud can be safely neglected, and in the domain of validity of the pseudo-potential approximation. The estimates for the temperature and size of the ion cloud corresponds to results of a simulation whose reliability was verified within better than 30% accuracy up to Mathieu parameters of the order of 0.5. Cooling and evaporation characteristic times estimates are shown to yield satisfactory predictions for the same range of Mathieu parameters. The condition for the thermal equilibration of the ion cloud involves a ratio of both characteristic times, which primarily depends on the so-called $\alpha_{Er} = \frac{qD_{reff}}{kT_{eff}}$ ratio. General considerations are given on the conditions in which space charge effects, not accounted for in the model, can be safely neglected. The formulae derived in the framework in this model can therefore be used as rule of thumbs for designing future devices, linear or 3D Paul traps, where a control of the ion cloud properties is required.

## 6. Acknowledgements

The author would like to thank O. Juillet and P. Van Isacker for their kind assistance.